\documentclass{ws-ijmpe}
\usepackage{units}

\begin{document}

\markboth{W. Sommer, C. Blume, J.F. Grosse-Oetringhaus, F. Kramer}{Quarkonia Measurements with the Central Detectors of ALICE}

\catchline{}{}{}{}{}
%%%%%%%%%%%%%%%%%%%%%%%%%%%%%%%%%%%%%%%%%%%%%%%%%%%%%%%%%%%%%%%%%%%%

\title{Quarkonia measurements with the central detectors of ALICE
%\footnote{For the title, try not to use more than 3 lines. Typeset the title in 10 pt Times roman, uppercase and boldface.}
}

\author{\footnotesize{Wolfgang Sommer, Christoph Blume, Frederick Kramer}}
%\footnote{wolfgang.sommer@ikf.uni-frankfurt.de}}

\address{
Institut f{\"u}r Kernphysik, University of Frankfurt,\\
Max-von-Laue Str. 1, Frankfurt, 60438, Germany
%\footnote{State completely without abbreviations, the affiliation and mailing address, including country. Typeset in 8~pt Times italic.}
\\
wolfgang.sommer@ikf.uni-frankfurt.de
}

\author{\footnotesize{Jan Fiete Grosse-Oetringhaus}}

\address{Institut f{\"u}r Kernphysik, University of M{\"u}nster,\\
Wilhelm-Klemm-Str. 9, M{\"u}nster, 48149, Germany}

\author{\footnotesize{for the ALICE collaboration}}

\maketitle

\begin{history}
\received{(received date)}
\revised{(revised date)}
%\accepted{(Day Month Year)}
%\comby{(xxxxxxxxxx)}
\end{history}

\begin{abstract}
A Large Ion Collider Experiment -- ALICE will become operational with the startup of the Large Hadron Collider -- LHC at the end of 2007. One focus of the physics program is the measurement of quarkonia in proton-proton and lead-lead collisions. Quarkonia states will be measured in two kinematic regions and channels: di-muonic decays will be measured in the forward region by the muon arm, the central part of the detector will measure di-electronic decays. The presented studies show the expected performance of the di-electron measurement in proton-proton and central lead-lead collisions. 
\end{abstract}

\section{Introduction}
Already 1985, the suppression of quarkonia in heavy ion collisions relative to proton-proton collisions was suggested as a sign for deconfined matter that is produced in these collisions\cite{Satz}. Over the last two decades various experiments\cite{NA50} \cite{NA60} \cite{Phenix} measured a significant suppression signal. Although strong indications for a J/$\Psi$ suppression were found the theoretical picture remains unclear. The startup of the LHC with the unprecedented center of mass energy of \unit[5.5]{TeV} per nucleon in lead-lead collisions the deconfined phase is expected to be larger, hotter and longer lived. Thus one expects new insights into the nature of quarkonia suppression.

\section{Quarkonia production at LHC energies}
To estimate the quarkonia production cross section, predictions from the Color Evaporation Model (CEM)\cite{CEMPPR} \cite{AlicePPR} were used. The cross sections are calculated as a product of the $q \bar q$ cross section and a transition probability to the individual states. A comparison to measured cross sections at the highest available Tevatron energy was done and the calculations were tuned to reproduce experimental data. The scaling from proton-proton to lead-lead collisions was done using the number of binary collisions, taking into account nuclear modifications employing the EKS98\cite{EKS98} parametrization. The resulting cross sections are shown in Table \ref{cross-sections}.
\begin{table}
\caption{Expected cross sections for quarkonia at the LHC. The cross sections include the branching into di-electrons as well as feed down from higher states. All numbers are given in $\mu$b{\protect \cite{AlicePPR}}.}
\begin{center}
\label{cross-sections}
\begin{tabular}[t]{|c|c|c|c|c|c|c|c|c|}
\hline
System&$\sqrt{s}$&$J/\Psi$&$\Psi$`&$\Upsilon$&$\Upsilon$`&$\Upsilon$``\\ \hline
proton-proton &\unit[14]{TeV}&3.18&0.057&0.02&0.005&0.003\\ 
proton-proton &\unit[5.5]{TeV}&1.83&0.033&0.009&0.002&0.0013\\
lead-lead &\unit[5.5]{TeV}&48930&879&304&78.8&44.4 \\ \hline
\end{tabular}
\end{center}
\end{table}

\section{The ALICE detector}
ALICE will be the only dedicated heavy-ion experiment at the LHC. It was optimized to measure a large variety of heavy-ion observables at a centre of mass energy of \unit[5.5]{TeV} per nucleon. Besides others, the measurement of quarkonia is a key element of the physics program. Quarkonia decaying into di-muons will be measured in the muon arm covering forward rapidities (-4$<$y$<$2.5)\cite{Martinez}. Di-electronic decays will be measured in the central barrel ($|$y$|<$0.9) involving the following three detectors:\\

\vspace{-8pt}
\noindent{\bf Inner Tracking System (ITS)\\}
The Inner Tracking System consists of three different types of silicon detectors, a silicon pixel detector (SPD), a silicon strip detector (SSD) and a silicon drift detector (SDD) arranged in six layers around the beam axis. 
The resolution of the combined ITS is $\sim$\unit[70]{$\mu$m} in r$\phi$ and $\sim$\unit[170]{$\mu$m} in z at p$_{t}$=\unit[1]{GeV/c} and $\sim$\unit[20]{$\mu$m} (\unit[60]{$\mu$m}) at p$_{t}$=\unit[10]{GeV/c}. Thus it will enable the measurement of open charm and beauty decays via their displaced vertices.\\

\vspace{-8pt}
\noindent{\bf Time Projection Chamber (TPC)}\\
With a diameter of $\sim$\unit[5]{m}, a length of $\sim$\unit[5]{m} and a volume of more than \unit[95]{m$^{3}$} the ALICE Time Projection Chamber is the largest TPC ever built. It is the main tracking device with a momentum resolution of less than 2.5\% for \unit[4]{GeV/c} electron tracks. Besides that it will also serve as a particle identification detector using the specific energy loss of particles in the TPC gas (90\% Ne, 10\% CO$_{2}$). \\

\vspace{-8pt}
\noindent{\bf Transition Radiation Detector (TRD)}\\
The Transition Radiation Detector was added to the experiment to improve the tracking capabilities and to discriminate electrons from the huge background of pions and other hadrons. The TRD will also serve as a fast trigger device for high p$_{t}$ electrons. It cylindrically surrounds the TPC with an inner diameter of $\sim$\unit[6]{m} and an outer diameter of $\sim$\unit[6.8]{m} and is segmented into 18 {\it{super modules}}. Each of these super modules is segmented in 5 {\it{stacks}} in z-direction with 6 layers of readout chambers in radial direction per stack. This gives a total of 540 read out chambers covering an area of \unit[736]{m$^{2}$} with 1.16 million readout channels.
\section{Simulation technique}
All presented studies were done within the ALICE offline software framework {\it{aliroot}}\cite{aliroot}. The aim was to simulate the expected performance of the di-electron quarkonia measurement with realistic estimates on the quarkonia production ratio, background and detector response. During one ALICE running year we expect to record 2$\cdot$10$^{9}$ proton-proton collisions and 2$\cdot$10$^{8}$ central lead-lead collisions. The complete simulation and reconstruction of one lead-lead event ($dN/dy_{ch}=3000$) takes about 5 hours on a modern computer. It is clear that this amount of events exceeds the available computing facilities by far. For this reason a fast simulation framework was developed. Instead of using the time consuming full detector simulation a look up table was created\cite{Grosse}, parameterizing the detector efficiency and resolution with respect to the kinematic variables p$_{t}$, $\theta$ and $\phi$ as well as for particle identification. 
\begin{figure}[b]
\begin{minipage}{0.48\textwidth}
\begin{center}
\epsfig{file=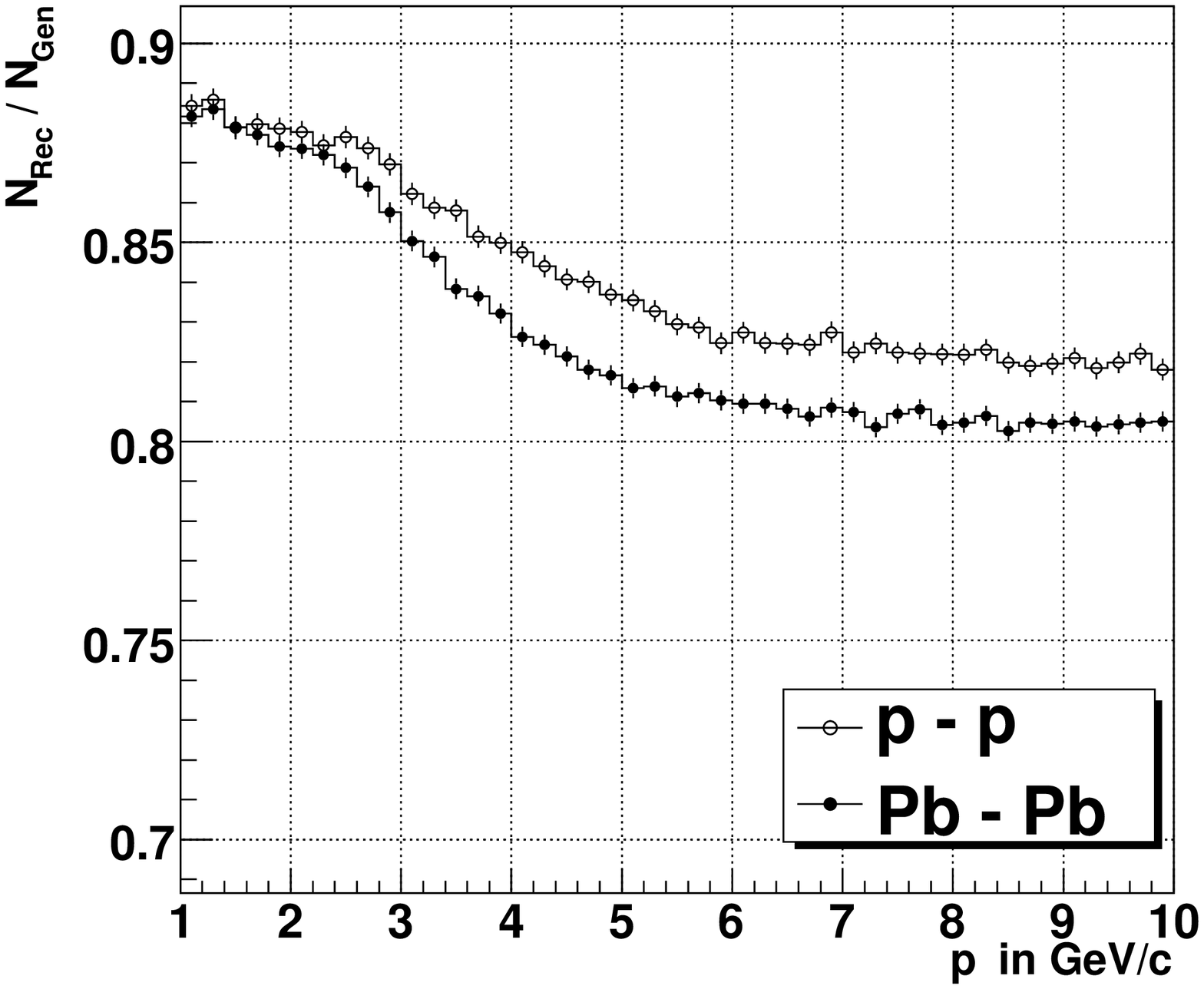,width=\textwidth}
\end{center}
\vspace*{-8pt}
\caption{Response function for tracking efficiency in p$_{t}$. The efficiency for high p$_{t}$ electrons decreases with p$_{t}$ since more straight tracks tend to stay in dead zones while curved tracks have a higher probability to re-enter the detector.}
\label{fig.response_pt}
\end{minipage}% Dies Prozent ist wichtig! (kein horiz. Abst. zw. minipages)
\begin{minipage}{0.04\textwidth}
\hfill % Damit die getrennte Beschriftung auch Abstand hat
\end{minipage}%
\begin{minipage}{0.48\textwidth}
\begin{center}
\epsfig{file=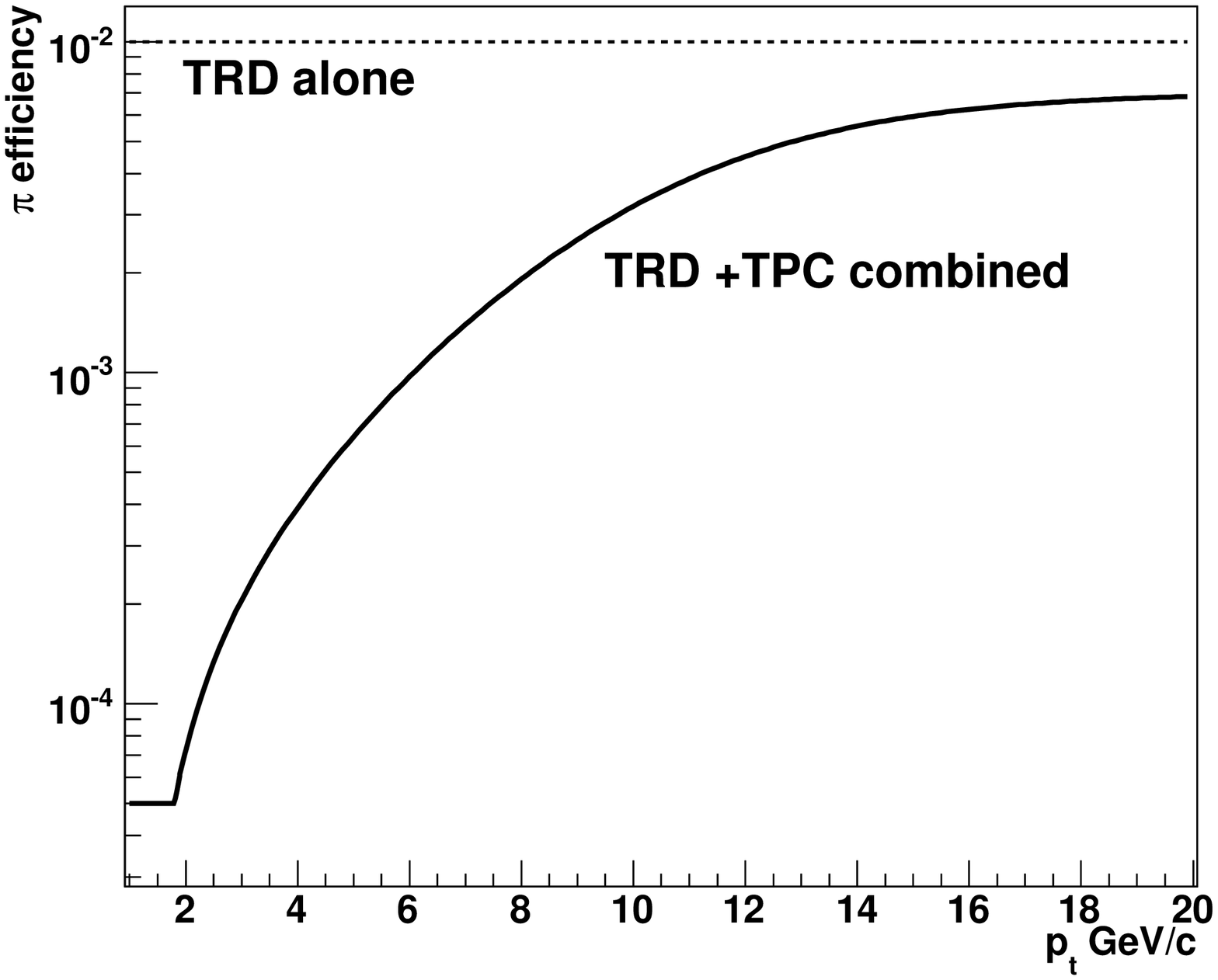,width=\textwidth}
\end{center}
\vspace*{-8pt}
\caption{The parameterization for the particle identification shows the used pion efficiency for 90\% electron efficiency (TRD alone) and 81\% electron efficiency for the combination of TRD and TPC.}
\label{fig.response_pid}
\end{minipage}
\vspace*{-24pt}
\end{figure}
Figure \ref{fig.response_pt} and  \ref{fig.response_pid} show examples of the parameterization. The look up table was created by analysing 5000 simulated Hijing-events with embedded electrons. \\
To generate the input to the fast simulations a cocktail of particles was created. 
The output of this generator was evaluated using the detector parameterization and analysed afterwards. 

\section{Results}
\subsection{Proton-proton collisions}

\begin{figure}[h]
  \begin{minipage}{0.56\textwidth}
    \begin{center}
      \epsfig{file=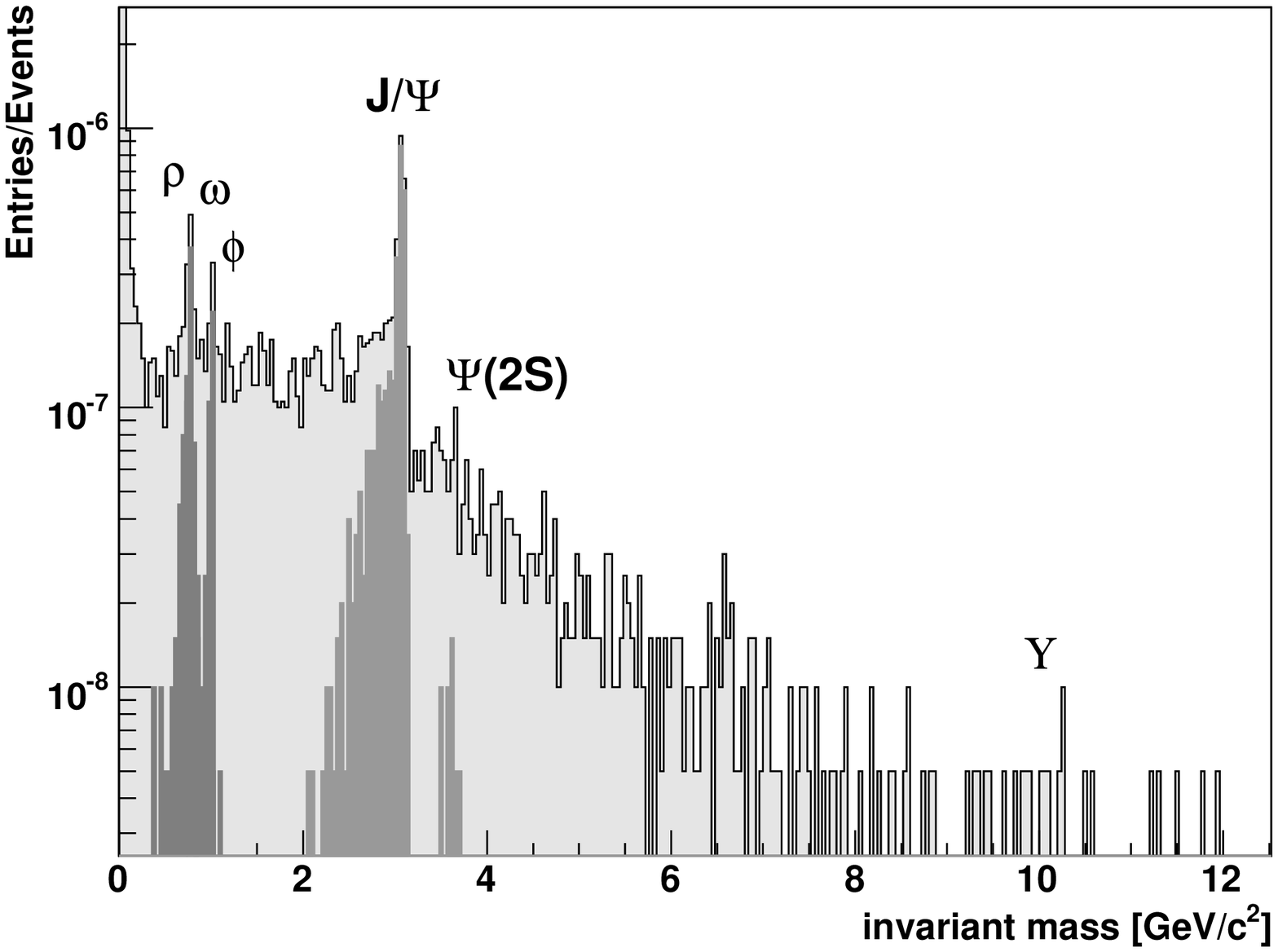,width=0.9\textwidth}
    \end{center}
  \end{minipage}% Dies Prozent ist wichtig! (kein horiz. Abst. zw. minipages)
  \begin{minipage}{0.04\textwidth}
     \hfill % Damit die getrennte Beschriftung auch Abstand hat
  \end{minipage}%
  \begin{minipage}{0.44\textwidth}
    \begin{center}
      \epsfig{file=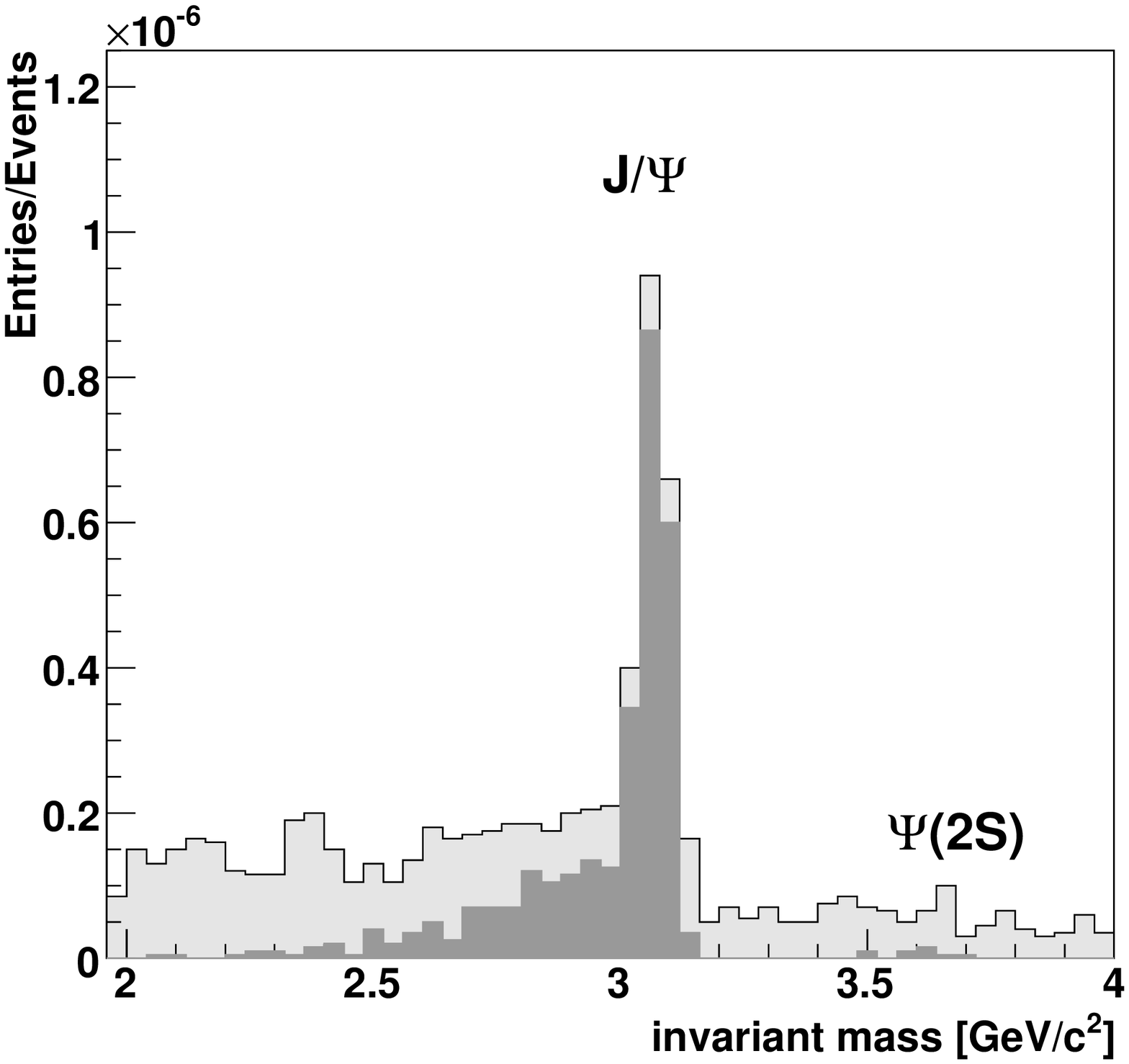,width=0.9\textwidth}
     	 \end{center}
  \end{minipage}
  \caption{The invariant mass distribution as obtained by the described simulation of 2$\cdot$10$^{8}$ minimum bias proton-proton events. To the left the complete spectrum is shown while the right plot focuses on the J/$\Psi$ region. The J/$\Psi$ peak is clearly visible with 360 entries and a Gaussian width of \unit[30]{MeV/c$^{2}$}. The signal to background ratio is 9 and the significance is 18. Apart from the J/$\Psi$ no other quarkonia state gives a significant signal. However, one should note that a trigger on di-electrons will significantly enhance the expected yields.}
\end{figure}

For the evaluation of the proton-proton measurement performance 2$\cdot$10$^{8}$ minimum bias events were generated\cite{Kramer}. This represents 20\% of the expected events to be recorded during one proton-proton run and enough statistics to extrapolate the results to full statistics. The events were generated using a cocktail of quarkonia simulated with the rates given in Table \ref{cross-sections} and a proton-proton PYTHIA\cite{Pythia} event simulating the background including open charm and beauty. A single electron p$_{t}$ cut of \unit[1]{GeV/c} was applied. One finds a significant J/$\Psi$ peak. Fitting this peak with a Gaussian gives 360 entries within the area of $\pm 1.5\sigma$ around the mean value and a width of \unit[30]{MeV/c$^{2}$}. The like-sign distribution was used to estimate the background. Since like-sign pairs are produced with much less probability than unlike-sign pairs in proton-proton collisions the distribution had to be scaled to match the unlike-sign distribution. After scaling the distribution it was used for background estimations. 
Table \ref{results} summarizes the obtained results. \\
One should note that the yield of measured quarkonia states will be substantially enhanced by the usage of a level one trigger on electrons. Assuming a conservative trigger efficiency of 10\% and a design luminosity for \unit[14]{TeV/c$^{2}$} proton-proton collisions of \unit[5$\cdot$10$^{30}$]{cm$^{-2}$s$^{-1}$} one would expect 9$\cdot$10$^{5}$ J/$\Psi$ and 1.4$\cdot$10$^{3}$ $\Upsilon$ to be recorded.

\subsection{Lead-lead collisions}
\begin{figure}[th]
\centerline{\psfig{file=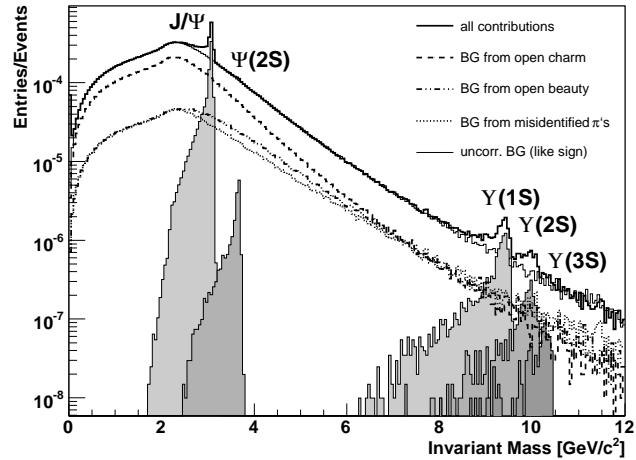,width=8.5cm}}
\caption{Invariant mass distribution as expected for 2$\cdot$10$^{8}$ central lead-lead collisions. A single electron p$_{t}$ cut of \unit[1]{GeV/c} was applied. The solid black line shows the unlike-sign distribution. The background was estimated using like-sign pairs (thin black line). The gray areas indicate the pure quarkonia distributions. The dashed lines show the various contributions from other sources to the distributions.}
\label{IMPbPbfull}
\end{figure}

\begin{figure}[h]
\begin{minipage}{0.48\textwidth}
\begin{center}
\epsfig{file=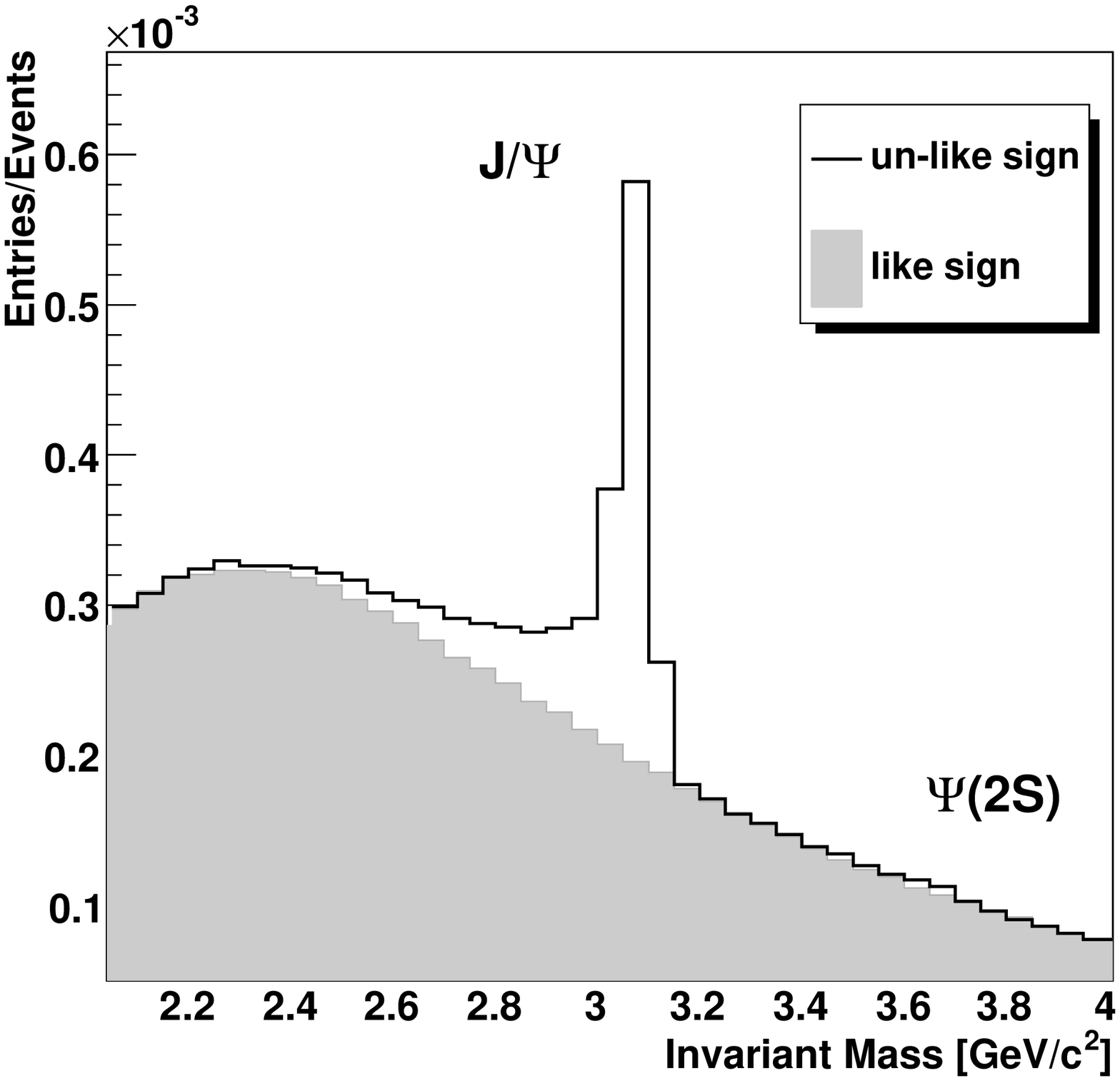,width=0.9\textwidth}
\end{center}
\end{minipage}% Dies Prozent ist wichtig! (kein horiz. Abst. zw. minipages)
\begin{minipage}{0.04\textwidth}
\hfill % Damit die getrennte Beschriftung auch Abstand hat
\end{minipage}%
\begin{minipage}{0.48\textwidth}
\begin{center}
\epsfig{file=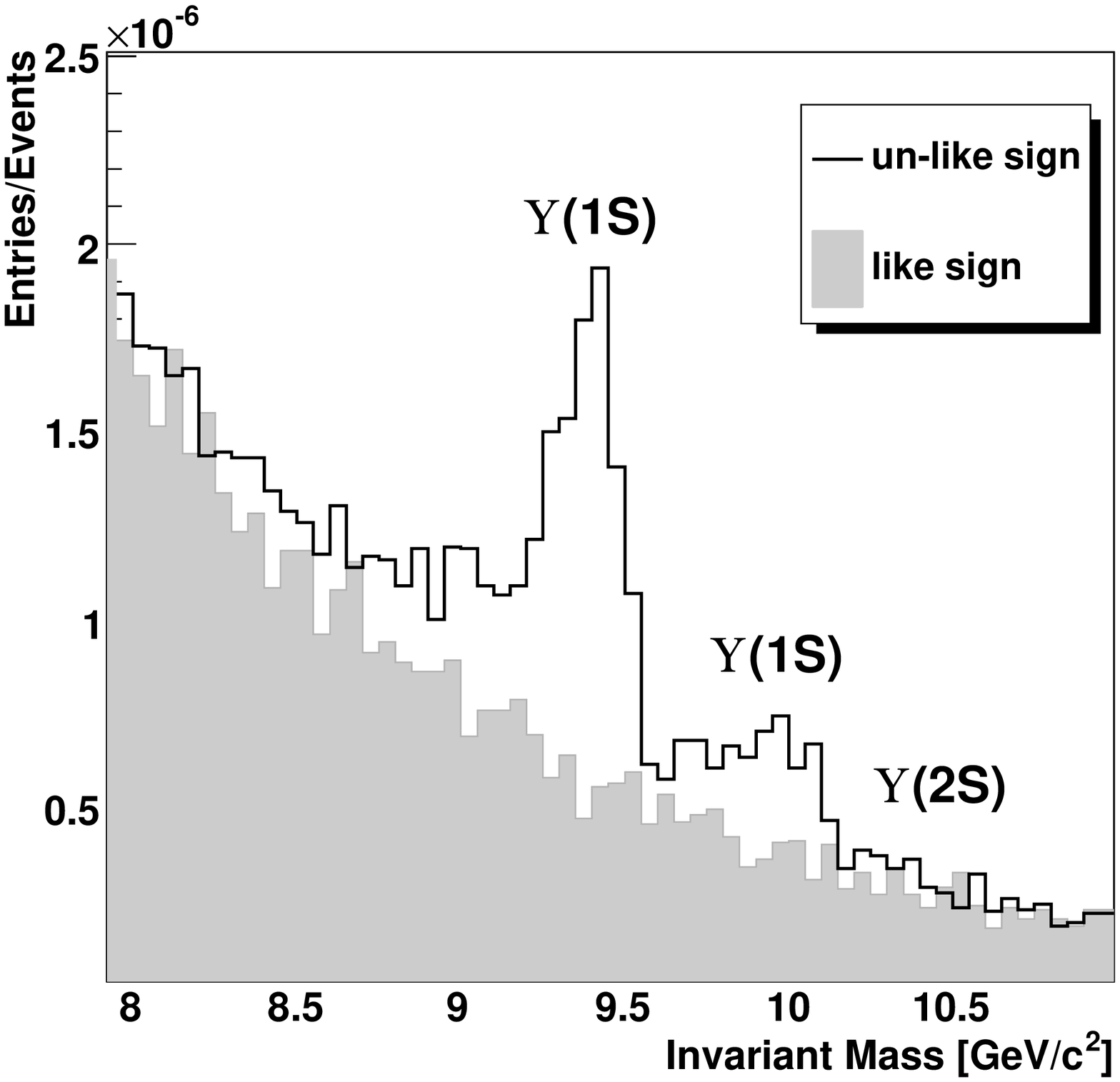,width=0.9\textwidth}
\end{center}
\end{minipage}
\caption{The invariant mass spectrum in the J/$\Psi$ (left) and $\Upsilon$ (right) mass region. The signal to background ratio for the J/$\Psi$ is 1.2, for the $\Upsilon$(1S) 1.1 and for the $\Upsilon$(2S) 0.8. $\Psi$(2S) and the $\Upsilon$(3S) do not have a significant peak. A fit of a Gaussian to the observed peaks gives a width of \unit[30]{MeV/c$^{2}$} for the J/$\Psi$ and \unit[90]{MeV/c$^{2}$} for the $\Upsilon$.}
\end{figure}
Figure \ref{IMPbPbfull} shows the result for lead-lead collisions. 2$\cdot$10$^{8}$ central events were simulated representing the expected statistics for one ALICE running year\cite{AlicePPR}. The event cocktail consisted of quarkonia signals which were simulated given the rates in Table \ref{cross-sections}. To account for electrons coming from semi-leptonic charm and beauty decays 100 $c \bar c$ and 4 $b \bar b$ per event were embedded. Pionic background was simulated using a parameterized HIJING generator tuned to the multiplicity of dN$_{ch}$/dy=3000. There are significant signals for the J/$\Psi$,  $\Upsilon$(1S) and $\Upsilon$(2S). $\Psi$(2S) and $\Upsilon$(3S) have only small contributions above background. The background was estimated using like-sign technique. After subtraction of the background and fitting the peaks with a Gaussian one obtains a width of \unit[30]{MeV/c$^{2}$} for the J/$\Psi$ and \unit[90]{MeV/c$^{2}$} for the $\Upsilon$, this resolution is sufficient to resolve the individual $\Upsilon$ states. All further results are summarized in Table \ref{results}.

\begin{table}
\begin{center}
\caption{Summary of the main results for minimum bias proton-proton and central lead-lead collisions. Values are only given for quarkonia states where significant signals are expected. The proton-proton results are scaled to 1$\cdot$10$^9$ events (one ALICE running year). See text for detailed description.}
\vspace*{-8pt}
\label{results}
\begin{tabular}[t]{|c|c|c|c|c|c|c|}
\hline

System								& 				&$J/\Psi$	&$\Upsilon$(1S)	&$\Upsilon$(2S)\\ \hline
             								& signals			& 1800 	& 15		& -- \\
\raisebox{1.5ex}[-1.5ex]{proton-proton} 		& $S/B$        		& 9 		& -- 		& -- \\
\raisebox{1.5ex}[-1.5ex]{$\sqrt{s}$=14 TeV}     	& $S/\sqrt{S+B}$     	& 40 		& -- 		& -- \\ \hline
                   							& signals 			& 120.000	& 900 	& 350 \\       
\raisebox{1.5ex}[-1.5ex]{lead-lead} 			& S/B 			& 1.2 	& 1.1		& 0.35 \\ 
\raisebox{1.5ex}[-1.5ex]{$\sqrt{s}$=5.5 TeV}    	& $S/\sqrt{S+B}$	& 245 	& 21 		& 8 \\ \hline

\end{tabular}
\end{center}
\end{table}

\vspace*{-2pt}
\section{Summary}
The expected performance of the quarkonia measurement in the di-electron channel has been simulated under realistic assumptions. Within one year of running ALICE will accumulate enough statistics to measure J/$\Psi$s in proton-proton and J/$\Psi$s, $\Upsilon$(1S) and $\Upsilon$(2S) in lead-lead collisions with good signal to background ratio and significance. The invariant mass resolution is sufficient to resolve the individual $\Upsilon$ states.
\vspace*{-2pt}
\section*{Acknowledgements}
This work was supported by the German Federal Ministry of Education and Research.

\end{document}